# Apport de l'imagerie cérébrale pour comprendre les déficits linguistiques

Charlotte Jacquemot[1-2] et Marine Lunven[1-2]

1. Département d'Etudes Cognitives, École normale supérieure, PSL University, 75005 Paris, France
2. University Paris Est Creteil, INSERM U955, Institut Mondor de Recherche Biomédicale, Equipe NeuroPsychologie Interventionnelle, F-94010 Creteil, France

La psycholinguistique et la neurolinguistique sont deux disciplines complémentaires qui étudient le langage sous des angles différents. La psycholinguistique s'intéresse aux processus cognitifs impliqués dans la production et la compréhension du langage, tandis que la neurolinguistique examine les bases cérébrales de ces mécanismes. Les techniques d'imagerie cérébrale permettent d'identifier les régions et réseaux impliqués dans les différents processus psycholinguistiques, et de mieux comprendre les effets de lésions cérébrales sur les capacités linguistiques. En intégrant les approches psycholinguistiques et neurolinguistiques, la recherche permet de mieux comprendre le traitement du langage, qu'il soit intact ou altéré. Ses connaissances ouvrent ainsi la voie à des stratégies de rééducation plus ciblées et efficaces.

## 1 Cerveau

### 1.1 Structure et composition

Le cerveau est un organe du système nerveux central du corps humain. Il est impliqué dans le traitement, la régulation et la coordination des informations et activités sensorielles, motrices, cognitives et comportementales.

#### 1.1.1 Protections du cerveau

Le cerveau est protégé de l'extérieur grâce à la boite crânienne (structure osseuse) et par les méninges, des couches de tissus protecteurs qui enveloppent le cerveau. Il est protégé de l'intérieur grâce à la barrière hémato-encéphalique qui filtre les substances présentes dans le sang et offre une protection relative du système nerveux central contre les infections (bactéries, virus) et les substances toxiques qui pourraient être présentes dans le sang.

Le cerveau comprend quatre ventricules, un ensemble de structures interconnectées impliquées dans la sécrétion du liquide cérébrospinal ou liquide céphalorachidien. Ce liquide circule autour et à l'intérieur du système nerveux central et a trois principales fonctions : une fonction de protection mécanique, une fonction de nutrition et d'évacuation des déchets, et une fonction de protection immunitaire. Le liquide offre une protection mécanique, il absorbe et amortit les mouvements ou les chocs contre les traumatismes extérieurs qui risqueraient d'endommager le cerveau. Il permet également de diminuer l'effet de la gravité sur le cerveau et de réduire son poids apparent de 97%, ce qui réduit la pression exercée par le cerveau sur les structures nerveuses et les vaisseaux sanguins, grâce à l'effet de flottabilité. Le liquide céphalorachidien permet la circulation des nutriments, des neurotransmetteurs et des facteurs de l'immunité. Il est renouvelé plusieurs fois par jour : il est sécrété et résorbé sans arrêt et est entièrement changé 3 à 4 fois en 24h.

### 1.1.2 Substance grise

Le cerveau humain fonctionne grâce à un réseau complexe de neurones, qui sont les cellules nerveuses responsables de la transmission des informations. Les neurones, dont le nombre est estimé entre 86 et 100 milliards, sont composés d'un corps cellulaire – le cœur de la cellule neuronale –, et de deux types de prolongements : les dendrites et un axone. Les dendrites sont habituellement plus courtes que l'axone et reçoivent l'influx nerveux des autres neurones. Elles forment avec le corps cellulaire, la substance grise ou cortex cérébral qui est la couche la plus externe du cerveau, la couche corticale. L'axone est un long prolongement qui va transmettre l'influx nerveux à d'autres cellules. Les neurones communiquent entre eux via des synapses, utilisant des neurotransmetteurs chimiques qui vont créer un influx nerveux.

### 1.1.3 Substance blanche : la glie

En plus des neurones, les cellules gliales tout aussi importantes et environ trois fois plus nombreuses que les neurones, composent le cerveau. Les cellules gliales comprennent les oligodendrocytes, les astrocytes et la microglie. Les oligodendrocytes fabriquent la myéline, une gaine protectrice présente le long des axones des neurones qui permet d'accélérer la transmission de l'influx nerveux entre les cellules. Les astrocytes et la microglie jouent un rôle de soutien, un rôle nutritif et un rôle de protection immunitaire. Les cellules gliales sont aussi appelées substance blanche. La substance blanche se trouve principalement sous la surface corticale, reliant les différentes parties du cerveau et de la moelle épinière par des faisceaux d'axones myélinisés.

### 1.1.4 Deux hémisphères et quatre lobes par hémisphère

Le cerveau est composé de deux hémisphères, le droit et le gauche qui sont globalement symétriques. Ils sont reliés par un faisceau de fibres de substance blanche appelé le corps calleux responsable de la transmission d'informations entre les deux hémisphères. Chaque hémisphère est divisé en quatre aires corticales ou lobes : lobe frontal, lobe pariétal, lobe occipital et lobe temporal. Les lobes frontal et pariétal sont séparés par le sillon central ou de Rolando, et les lobes frontal et temporal par le sillon latéral ou de Sylvius (Figure 1).

Les lobes cérébraux ainsi que le cervelet sont formés de replis avec des circonvolutions ou gyrus, et des sillons ou sulcus. Ces replis permettent d'augmenter grandement la surface cérébrale, tout en gardant un volume restreint et une boîte crânienne de taille raisonnable et d'un volume de 1,2 à 1,5 litre. Les principaux replis se retrouvent chez tous les individus, cependant la localisation précise varie en fonction des individus. Il existe une spécialisation de fonction au sein des régions, par exemple la région postérieure du cerveau où se situent les lobes occipitaux est impliquée dans le traitement des informations visuelles. En dessous du lobe occipital et en arrière du tronc cérébral se situe le cervelet. Cette structure est impliquée dans le contrôle de l'équilibre et la coordination des mouvements. Le tronc cérébral sert de point de passage entre les hémisphères cérébraux et la moelle épinière. Le tronc cérébral, le cervelet et le cerveau forment l'encéphale.

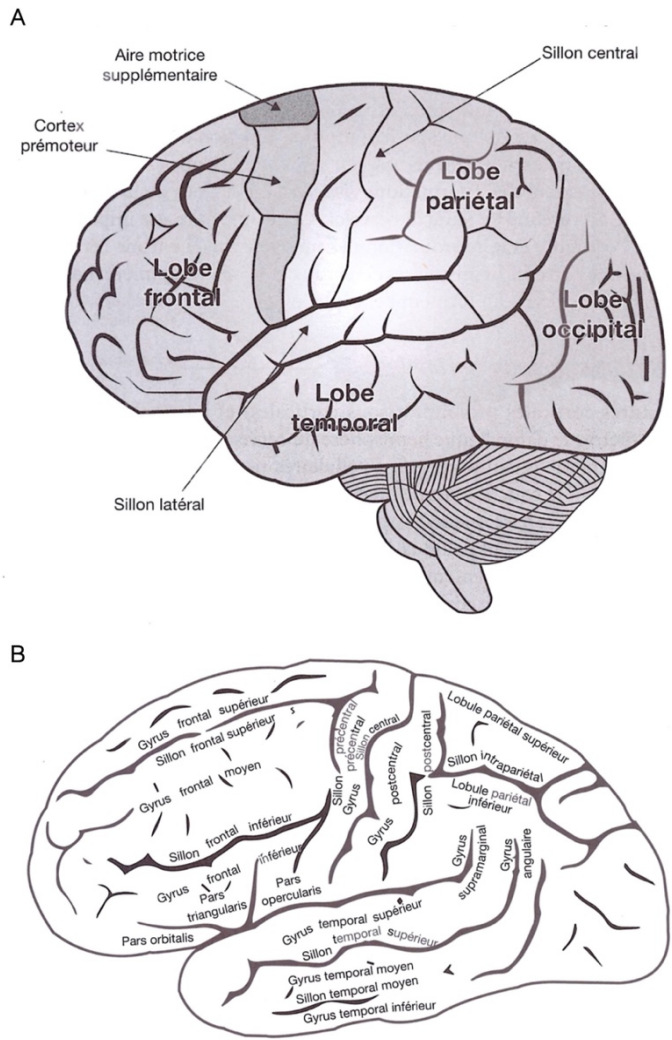

*Figure 1. Hémisphère gauche du cerveau et ses différentes régions cérébrales (à gauche l'avant). (A) lobes et (B) sillons et gyri.*

1.1.5   Structures sous-corticales

Des structures corticales profondes (sous-corticales) et bilatérales sont situées de manière symétrique dans chaque hémisphère du cerveau (Figure 2). Elles sont composées d'une grande quantité de corps cellulaires neuronaux et de dendrites, tout comme le cortex – la couche externe du cerveau – et forment de la matière grise.

Les ganglions de la base ou noyaux gris centraux comprennent le striatum (noyau caudé et putamen), le globus pallidus, le noyau sous-thalamique et la substance noire. Ils sont impliqués principalement dans le contrôle et la coordination motrice, ainsi que dans divers aspects du comportement et de la cognition notamment, mais aussi l'apprentissage, la motivation, et la prise de décision.

Le thalamus est une structure centrale et multifonctionnelle du cerveau, agissant comme un relai et un centre de modulation pour les informations sensorielles, motrices et cognitives.

En raison de ses nombreuses connexions avec d'autres régions du cerveau, il joue un rôle crucial dans l'intégration, la transmission et la modulation des informations entre les différentes parties du cerveau. Il intervient également dans la régulation de l'éveil, de l'attention et de la conscience. Les connexions thalamo-corticales sont essentielles pour le maintien de l'état d'éveil et la modulation des cycles veille-sommeil.

L'amygdale est une structure en forme d'amande située profondément dans les lobes temporaux de chaque hémisphère cérébral. C'est une structure cérébrale essentielle pour la régulation des émotions, la mémoire émotionnelle et la réponse au stress. Elle est fortement impliquée dans les réponses autonomes et endocriniennes via des connexions avec l'hypothalamus. Elle est également connectée avec l'hippocampe.

L'hippocampe est une structure cérébrale du lobe temporal en arrière de l'amygdale impliquée dans la mémoire épisodique (voir glossaire) avec la formation et la récupération des souvenirs, la navigation spatiale et l'orientation, l'apprentissage et la régulation des émotions.

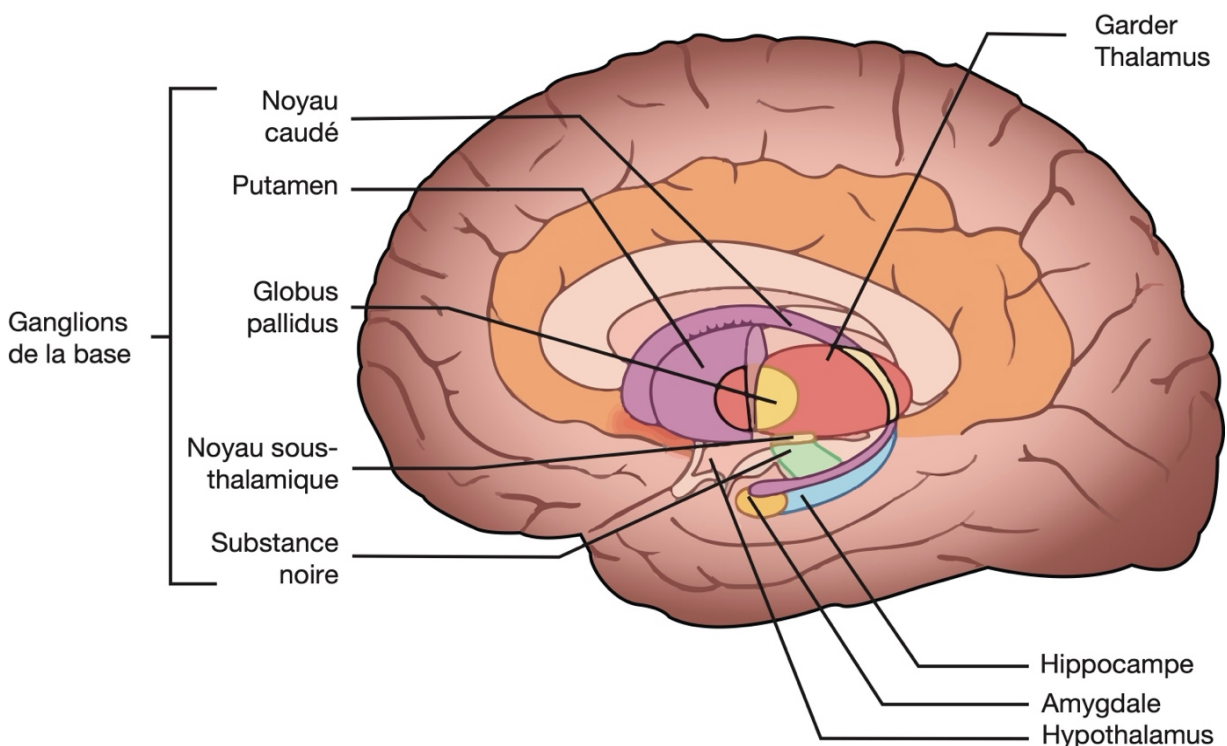

*Figure 2. Structures sous-corticales.*

## 1.2 Lésions cérébrales : les causes et les conséquences

Les lésions cérébrales acquises (par opposition aux anomalies développementales) peuvent avoir des causes multiples. En fonction des régions cérébrales touchées, les conséquences pour les personnes touchées seront différentes : les déficits peuvent être moteurs, sensoriels, et cognitifs et comportementaux (voir glossaire).

### 1.2.1 Accident vasculaire cérébral (AVC)

Les lésions cérébrales dont la cause est vasculaire touchent environ 150 000 personnes par an en France. Dans 80% des cas, l'AVC est ischémique : il est provoqué par l'obstruction d'un vaisseau sanguin par un caillot, réduisant le flux sanguin dans la zone du cerveau touchée. Dans 20% des cas, l'AVC est hémorragique : il est provoqué par la rupture d'un vaisseau sanguin, entraînant une hémorragie. Les trois quarts des personnes touchées par un AVC ont plus de 65 ans.

### 1.2.2 Traumatismes crâniens

Les lésions cérébrales traumatiques sont dues à un choc qui va impacter le tissu cérébral et en détruire une partie. En France, environ 150 000 à 170 000 personnes sont victimes de traumatismes crâniens par an. Les chocs peuvent être dus à des accidents de la voie publique (ex. voiture, moto, vélo) qui représentent la majorité des traumatismes crâniens et touchent en priorité les 15-30 ans, des chutes qui touchent majoritairement les enfants et les personnes âgées, des accidents de sport (ex. rugby, boxe, ski, football), et des violences physiques (ex. bébé secoué, violences domestiques).

### 1.2.3 Maladies neurodégénératives

Les maladies neurodégénératives sont caractérisées par une dégénérescence progressive et irréversible des cellules nerveuses dans une ou plusieurs régions du cerveau. En France, les maladies neurodégénératives sont diagnostiquées chez environ 255 000 personnes par an. La maladie d'Alzheimer, dont 225 000 nouveaux cas sont diagnostiqués chaque année, est caractérisée par la dégénérescence de plusieurs régions cérébrales, notamment l'hippocampe, le cortex entorhinal (région corticale située en dessous de l'hippocampe et impliquée dans les mécanismes d'olfaction et de la mémoire), et diverses parties du cortex cérébral. Dans la maladie de Parkinson et la maladie de Huntington, la dégénérescence cérébrale commence au niveau des ganglions de la base puis s'étend aux régions corticales au fur et à mesure que la maladie évolue. La sclérose en plaques est une maladie auto-immune chronique qui affecte le cerveau et la moelle épinière. Elle se caractérise par des lésions de démyélinisation, où la gaine de myéline qui entoure les fibres nerveuses est endommagée. Cela peut toucher différentes régions cérébrales qui dégénèrent suite à la démyélinisation.

### 1.2.4 Tumeurs cérébrales

En France, environ 5 000 à 6 000 nouveaux cas de tumeurs cérébrales, tant bénignes que malignes sont diagnostiqués par an. Les tumeurs peuvent comprimer ou détruire certaines structures cérébrales. En fonction de la structure cérébrale touchée, des déficits cognitifs, sensoriels, moteurs et comportementaux peuvent s'observer.

### 1.2.5 Infections

Certaines infections virales ou bactériennes vont traverser la barrière hémato-encéphalique et entrainer une infection cérébrale qui peut toucher certaines structures cérébrales. Les infections cérébrales représentent moins de 5000 cas par an en France. L'encéphalite est une inflammation du cerveau. La méningite est une inflammation des méninges.

### 1.2.6 Résection chirurgicale

La résection chirurgicale cérébrale est une chirurgie cérébrale qui consiste à enlever une zone du cerveau. Elle peut être réalisée pour traiter diverses pathologies neurologiques, notamment les tumeurs cérébrales, les épilepsies résistantes aux traitements, les malformations vasculaires et autres anomalies structurales. En fonction de la localisation de la zone qui est retirée, l'ablation d'une zone du cerveau peut avoir des conséquences sur le fonctionnement du cerveau et entrainer des déficits cognitifs.

### 1.2.7 Hypoxie ou anoxie cérébrale

Le manque d'oxygène dans le cerveau peut être causé par une noyade, une suffocation, une crise cardiaque, ou une intoxication au monoxyde de carbone. L'oxygène étant essentiel pour le fonctionnement des cellules, une privation prolongée peut entraîner des lésions cérébrales.

### 1.2.8 Intoxication

L'exposition à des substances toxiques, telles que les drogues, l'alcool, les métaux lourds (comme le plomb ou le mercure), le monoxyde de carbone, et certains produits chimiques peut impacter le fonctionnement normal des cellules cérébrales.

# 2 Langage

Le langage a un statut particulier, car c'est une fonction spécifique à l'espèce humaine, acquise sans apprentissage explicite par tous les enfants et qui peut être spécifiquement lésée chez des patient·es après une lésion cérébrale. La faculté de langage est une capacité de l'espèce humaine qui permet la rapide acquisition d'une ou des langues natives, et ce, peu importe les langues. Elle permet de générer à partir d'un nombre fini d'éléments (phonèmes), un nombre infini de phrases et de contenus. Les locuteurs et locutrices sont capables de produire et de comprendre des phrases qu'ils et elles n'ont jamais entendues.

« Est-ce que tu peux me passer le sel ? » cette question est apparemment très simple. Or, y répondre par un comportement adapté est beaucoup plus complexe qu'il n'y paraît. En effet, de nombreuses étapes sont nécessaires : la perception des sons, l'identification des mots, le traitement de la syntaxe de la phrase et la compréhension du sens de la question puis, l'interprétation du sens de cette question, au-delà d'une interprétation littérale, et enfin la décision de répondre à cette question non pas par l'affirmative ou la négative, mais par un geste envers la personne qui vous a sollicité·e, celui de lui tendre la salière. Tout cela nécessite non seulement de comprendre le langage, mais aussi de porter votre attention sur la question posée sans vous laisser distraire par votre téléphone qui sonne, de maintenir en mémoire cette question pendant que vous explorez vos connaissances conceptuelles et culturelles, et d'identifier le sens caché de cette phrase qui n'appelle pas une réponse binaire (oui ou non), mais un geste, une réponse motrice de votre part.

Initialement décrit comme un système bimodal, intégrant un système de production de la parole et un système de compréhension de la parole, le modèle de traitement du langage s'est largement complexifié depuis les travaux de Broca et Wernicke. En effet, en 1861, le docteur Broca, qui suivait le patient Leborgne depuis plusieurs années, fit le lien entre l'impossibilité pour ce patient à s'exprimer et une zone de la région frontale de l'hémisphère gauche du cerveau. L'observation d'un déficit de la parole corrélée à une lésion du gyrus frontal inférieur conduisit Broca à faire l'hypothèse que le centre de production de la parole était le lobe frontal inférieur, appelée aujourd'hui aire de Broca.

À la même époque, le neurologue Wernicke décrivit des troubles de compréhension de la parole chez des patient·es présentant des lésions à la jonction du gyrus temporal supérieur et du gyrus pariétal, toujours dans l'hémisphère gauche du cerveau. Deux centres du langage furent alors proposés, un centre de compréhension de la parole et un de production de la parole qui impliquaient respectivement l'aire de Wernicke et l'aire de Broca.

Un modèle qui – s'il y a le mérite d'être extrêmement simple – est loin de pouvoir expliquer les déficits que l'on observe chez les patient·es et loin de pouvoir rendre compte des activations neuronales liées au traitement du langage. Grâce à l'apport de la psycholinguistique et de la neurolinguistique, le modèle du langage s'est complexifié en intégrant plusieurs composantes aussi bien en compréhension qu'en production, tout en prenant en compte l'interaction avec d'autres systèmes, tels que le système de l'attention, la mémoire de travail (voir glossaire), le système conceptuel et le système de la prise de décision (voir glossaire).

## 2.1 Apport de la linguistique

La linguistique est l'étude des différentes composantes du langage. Même si le monde compte plusieurs centaines ou milliers de langues (l'ONU reconnaît 141 langues officielles et les estimations des linguistes décrivent de 3 000 et 7 000 langues vivantes), le langage repose sur des composantes que l'on retrouve dans toutes les langues du monde, qu'elles soient orales ou signées.

### 2.1.1 Phonétique

La phonétique décrit les propriétés acoustiques des sons de la parole, les traits phonétiques, et concerne la prononciation. Par exemple, en français la seule différence entre les sons « b » et « p » est que pour prononcer « b » les cordes vocales vibrent alors que pour prononcer « p » les cordes vocales ne vibrent pas, « b » est voisé contrairement à « p ». C'est le trait phonétique qui distingue ces deux consonnes.

### 2.1.2 Phonologie

La phonologie fait l'inventaire des sons ou phonèmes qui sont utilisés dans une langue donnée et décrit la façon dont ces sons sont agencés au sein des mots. Par exemple, « r » et « l » sont deux phonèmes du français alors qu'en japonais ces deux phonèmes n'en font qu'un. Par ailleurs, des règles phonologiques définissent comment les sons s'assemblent au sein des mots : une syllabe peut commencer par « pr», mais pas par « rp ». En revanche, une syllabe peut se terminer par « rp » comme dans harpe « arp ».

### 2.1.3 Morphologie

La morphologie décrit les morphèmes – les plus petites unités qui ont un sens – et la façon dont ces morphèmes sont agencés pour former des mots. Par exemple, « in » est un morphème qui permet d'exprimer le contraire, comme dans « infini », « impossible », « incorrigible ». Les formes masculine et féminine peuvent être marquées par des morphèmes comme dans les mots : directeur / directrice. Les formes du singulier et du pluriel peuvent être marquées par des morphèmes comme dans les mots : cheval / chevaux.

On parle de morpho-syntaxe pour désigner les accords entre les mots, par exemple entre le déterminant et le nom « la bouteille », entre le nom et l'adjectif « la fille brillante », entre le pronom et le verbe « nous sourions ».

### 2.1.4 Syntaxe

La syntaxe décrit les règles qui combinent les mots au sein des phrases. Par exemple, on dit « Octave marche dans la rue » et non pas « La rue dans Octave marche ». Pour que la phrase ait un sens, les mots sont dans un ordre défini par un arbre syntaxique qui permet de comprendre comment les mots de la phrase sont rattachés entre eux. Par exemple, dans la phrase « Ernest est entré quand il a souri », « Ernest » et « il » peuvent être soit la même personne ou soit deux personnes différentes. En revanche, dans la phrase « Il a souri quand Isidore est entré », « il » et « Isidore » sont obligatoirement deux personnes différentes. Cela est dû aux règles syntaxiques de co-référence du français. Les personnes qui ont appris le français comme langue native n'ont pas conscience de la grande majorité de ces règles et les appliquent implicitement sans savoir qu'elles existent.

### 2.1.5 Sémantique

La sémantique s'attache à la compréhension du sens. Le lexique décrit les mots du vocabulaire. Par exemple, « téléphoner » est un mot du lexique, mais « banoumer » n'en est pas un. Cependant comprendre le sens d'une phrase ne signifie pas additionner le sens des mots. La sémantique s'intéresse à la compréhension des phrases, ce qui impose de considérer chaque mot en fonction des autres mots de la phrase pour comprendre le sens global. Le même mot peut avoir différents sens en fonction des autres mots de la phrase. « Elle était vêtue d'une robe légère » signifie que la robe était d'un tissu fin et aéré, d'une apparence fluide et facile à porter. « Elle portait un sac léger » signifie que le sac était de petite dimension et ne pesait pas lourd, et « Son explication était un peu légère » signifie que l'explication manquait d'argumentation.

### 2.1.6 Prosodie

La prosodie concerne les caractéristiques rythmiques et mélodiques de la parole. Elle englobe les éléments suprasegmentaux, c'est-à-dire qui s'appliquent à des entités qui peuvent comprendre plusieurs segments ou phonèmes, tels que l'intonation, le rythme, l'accentuation et la durée des sons. L'intonation fait référence aux variations de hauteur (ou tonalité) de la voix. La prosodie linguistique peut indiquer des questions, des affirmations, des exclamations. Seule l'intonation diffère entre « Tu viens ! » et « Tu viens ? » alors que le sens est différent. Le rythme concerne la cadence et la vitesse de la parole, y compris les pauses et les segments de parole plus rapides ou plus lents. La prosodie linguistique peut aussi transmettre des informations qui ne sont pas explicitement exprimées par les mots, comme par exemple, un ton ironique qui peut modifier le sens de la phrase produite et signifier le contraire de ce qui est littéralement dans une phrase : « Cette réunion de famille est une véritable partie de plaisir » prononcée sur un ton ironique, signifie que la réunion est une corvée. La prosodie émotionnelle transmet des informations sur les émotions ressenties (ex. tristesse, peur, joie, surprise, colère, dégout).

### 2.1.7 Pragmatique

La pragmatique s'intéresse à la question de la transmission d'informations entre individus, en tenant compte non seulement du lexique, de la syntaxe, de la prosodie, mais aussi du contexte, des connaissances partagées entre les individus, et des intentions de la personne qui parle. La pragmatique renvoie aux informations qui sont transmises, mais qui ne sont pas présentes littéralement, ni explicites dans la phrase. Elle permet d'expliquer comment les individus vont pouvoir s'affranchir des ambiguïtés présentes dans les énoncés. Conjointement avec la prosodie, la pragmatique permet de transmettre des informations au-delà du discours, et fait aussi référence au second degré, à l'humour et au sens figuré.

## 2.2 Apport de la psycholinguistique à la compréhension des déficits langagiers

La psycholinguistique étudie les processus cognitifs du traitement du langage et explore comment les individus comprennent le langage et parlent. Depuis les descriptions de Broca et Wernicke, différents modèles qui rendent compte de la faculté de langage ont été développés. Un modèle fonctionnel récent du langage (Jacquemot et al., 2019) fait la synthèse de modèles de psycholinguistique développés sur la base de données recueillies chez des sujets sains et chez des sujets cérébrolésés présentant des troubles du langage (Caramazza, 1997 ; Dell & O'Seaghdha, 1992 ; Franklin, 1989 ; Friederici, 2011 ; Jacquemot et al., 2011 ; Levelt et al., 1999 ; Miceli et al., 1994 ; Ramoo et al., 2021). Ce modèle se concentre sur le langage oral et fait la distinction entre les modalités de compréhension, de production et de répétition de la parole. Chaque modalité est composée de plusieurs composantes qui reflètent les différentes composantes décrites en linguistique. Ces composantes ont été mises en évidence dans des études de psycholinguistique et peuvent être spécifiquement lésées chez les patient·es. Les déficits observés en fonction des différentes composantes lésées sont rapportés sur la *Figure 3*. En compréhension, le traitement phonologique est la transformation du signal acoustique de parole en phonèmes. Le traitement morphologique va identifier les différents morphèmes au sein des mots ou de la phrase. Le traitement lexical permet la segmentation du flux de parole en unité de sens et l'appariement des mots identifiés avec ceux stockés dans le dictionnaire mental, le lexique. Plusieurs mots possibles (candidats lexicaux) peuvent être activés simultanément, et le contexte et la prosodie vont aider à sélectionner le mot approprié. L'analyse syntaxique permet l'identification des rôles grammaticaux des mots (ex. sujet, verbe, objet) au sein de la phrase. Le traitement sémantique est l'interprétation de la phrase pour en extraire le sens global. Cette étape nécessite l'intégration des informations lexicales et syntaxiques avec les connaissances contextuelles et sémantiques (voir glossaire). Dans certains cas, cette étape nécessite également l'intégration d'informations prosodiques et pragmatiques pour comprendre le sens de l'énoncé. En effet, le sens global de l'énoncé est ajusté en fonction du contexte communicatif et des connaissances préalables de l'interlocuteur ou de l'interlocutrice. Cela inclut la compréhension des intentions communicatives, des sous-entendus et des éléments implicites. Pour la production de la parole, les mêmes étapes vont avoir lieu en sens inverse, du concept à l'articulation. L'intention de communication, l'information que l'on veut transmettre va générer à partir du système conceptuel, un arbre syntaxique dans lequel des items lexicaux vont être intégrés. L'étape morphémique permet de conjuguer et d'accorder en genre et en nombre les différents items lexicaux, en fonction de l'arbre syntaxique. Les phonèmes qui composent les différents mots sont récupérés et planifiés pour être articulés.

### 2.3 De la sémiologie à la psycholinguistique : le diagnostic fonctionnel

La sémiologie est l'étude des signes d'une pathologie, d'une maladie afin de réaliser un diagnostic et en fonction de ce diagnostic mettre en place un traitement adapté, c'est-à-dire une rééducation du langage adaptée au patient ou à la patiente.
La sémiologie classique en orthophonie consiste à classer les signes cliniques de manière binaire : aphasie fluente ou non fluente, compréhension préservée ou déficitaire, répétition préservée ou déficitaire et dénomination préservée ou déficitaire. L'aphasie de Broca est le résultat d'une aphasie non fluente, avec une compréhension préservée, une répétition déficitaire et une dénomination déficitaire. L'aphasie de Wernicke est le résultat d'une aphasie fluente voire d'une logorrhée qui peut être du jargon, avec une compréhension déficitaire, une répétition déficitaire et une dénomination plutôt préservée. Cependant, ce diagnostic en termes de profil clinique, s'il permet d'avoir un cadre commun de description des troubles du langage, ne permet pas de mettre en place une rééducation adaptée (Hillis, 1998). En effet, plusieurs travaux récents ont montré que la rééducation était plus efficace si elle était ciblée sur la composante linguistique déficitaire (Bachoud-Lévi et al., 2022 ; Hillis, 1993, 1998 ; Hillis & Caramazza, 1994 ; Jacquemot et al., 2012 ; Nickels, 2002). Dans le cadre d'une pratique clinique orthophonique, il est plus utile de déterminer que le déficit du patient ou de la patiente avec un trouble de la production de la parole concerne par exemple la composante phonologique afin de concentrer la rééducation sur cette composante lésée, plutôt que de diagnostiquer une aphasie de Broca dont le profil ne va pas permettre de mettre en place une rééducation de la composante lésée. Chaque composante langagière, si elle est lésée, entraine des troubles caractéristiques dont l'étude par les orthophonistes permet de préciser la composante lésée. Pour faire un diagnostic fonctionnel des différentes composantes du langage oral, plusieurs tests peuvent être utilisés dont la CALAP (Jacquemot et al., 2019), le MT 86 (Nespoulous et al., 1992), la BIA (Gatignol et al., 2012) et la BETL (Tran & Godefroy, 2011). La CALAP, outil d'évaluation rapide du langage oral, évalue les différentes composantes du modèle de traitement de langage de la

*Figure 3*.

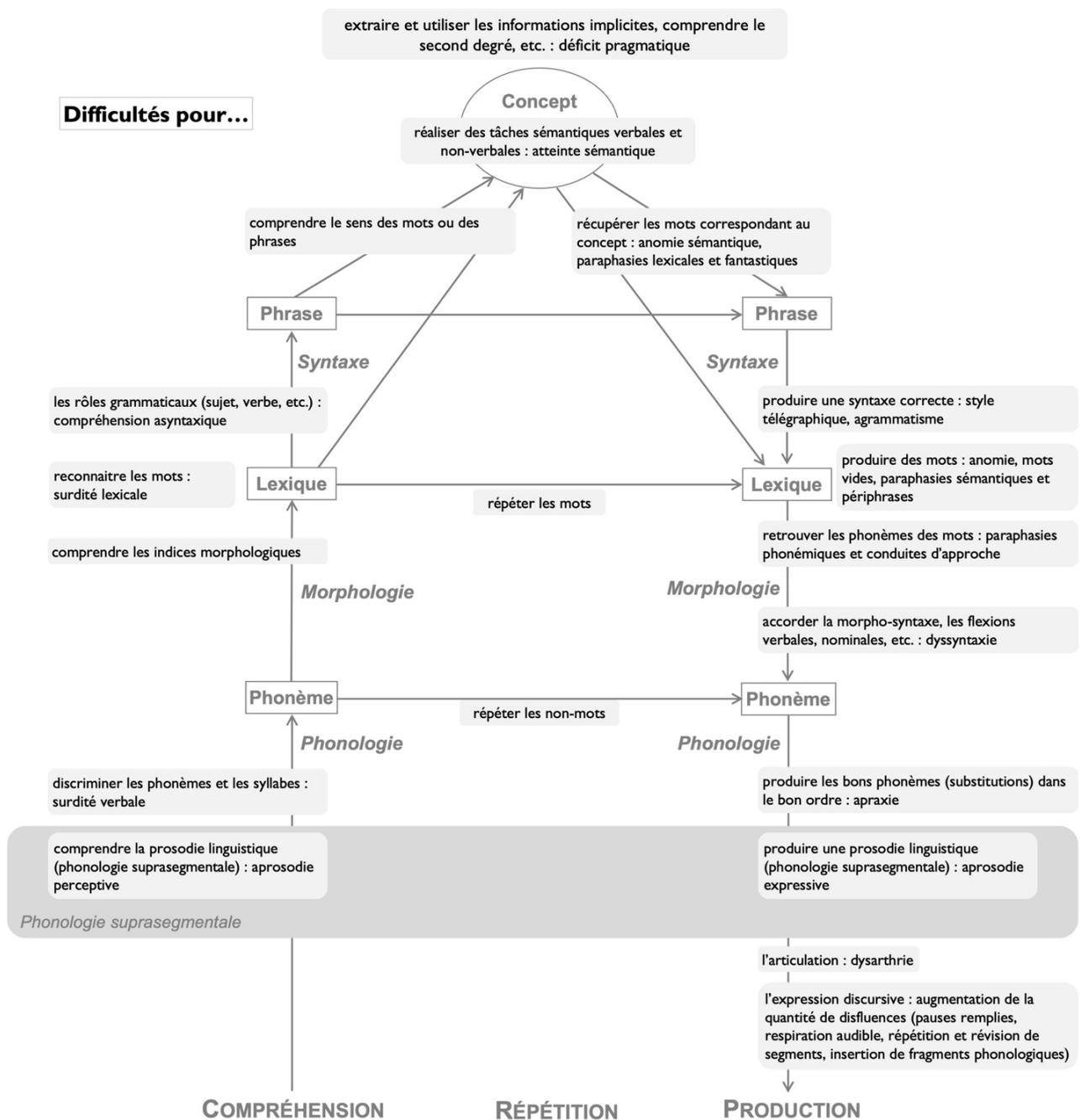

*Figure 3. Modèle de traitement du langage intégrant les processus de compréhension, de production et de répétition, et les difficultés langagières associées, en fonction des composantes lésées (Jacquemot et al., 2019). Voir le glossaire pour les définitions.*

Elle permet de faire un diagnostic direct de la composante langagière lésée et de mettre en place une rééducation ciblée et efficace (Bachoud et al., 2022). Les résultats aux autres tests nécessitent d'être analysés en se basant sur un modèle théorique du langage, pour identifier la composante lésée.

# 3   Neurolinguistique : faire le lien entre les troubles cliniques et les lésions cérébrales

Au 19e siècle, un tournant décisif dans la compréhension du traitement du langage a été initié par Broca et Wernicke mettant en lumière l'importance de l'hémisphère gauche pour comprendre et parler et l'idée que la production de la parole implique la région frontale inférieure, aire dite de Broca et que la compréhension de la parole implique une région à la jonction entre le lobe pariétal et temporal, dite aire de Wernicke. Suite à ces découvertes, au 20e siècle, les techniques d'imagerie cérébrale sont apparues, permettant de localiser précisément les lésions cérébrales. Avec le développement de l'imagerie fonctionnelle et son utilisation à grande échelle dans les années 1980, il est devenu possible d'identifier les régions activées pour telle ou telle activité. Les données acquises chez des participantes et participants sains ont permis de mettre en évidence l'activation de certaines régions cérébrales en lien avec des tâches réalisées. Cependant l'activation d'une région pour une tâche donnée ne garantit pas que cette région soit indispensable pour sa réalisation. C'est pourquoi l'apport des données de patient·es avec des lésions cérébrales a été primordial pour confirmer que les régions mises en évidence en imagerie cérébrale fonctionnelle étaient bien nécessaires aux fonctions étudiées.

## 3.1   Les méthodes d'imagerie cérébrale

### 3.1.1   Imagerie anatomique : l'Imagerie par Résonance Magnétique (IRM) et scanner

L'analyse des lésions cérébrales permet d'identifier les zones cérébrales endommagées par une lésion (ex. à la suite d'un AVC ou d'une tumeur) et de les associer aux déficits cognitifs observés. Cela s'obtient à partir d'images anatomiques acquises par IRM ou scanner. Les lésions sont délimitées manuellement sur les images du cerveau des patient·es, mis dans un référentiel standard (Figure 4A). Cela permet de visualiser en trois dimensions de façon précise l'étendue et la localisation des régions cérébrales endommagées. À partir de la délimitation des lésions, plusieurs types d'analyses peuvent être réalisés. De nombreuses études superposent les lésions des patient·es pour identifier les régions cérébrales les plus souvent affectées en cas de déficit (Figure 4B). La méthode de soustraction des lésions compare deux groupes : les patient·es avec un déficit et ceux sans, en superposant leurs lésions pour visualiser les zones cérébrales spécifiques à chaque groupe. Le voxel-lesion symptom mapping (VLSM, cartographie des symptômes en fonction des lésions voxelisées) permet d'aller plus loin en réalisant des analyses statistiques à partir des lésions reconstruites en trois dimensions. Par exemple, cette méthode inclut des analyses de régression pour déterminer les régions qui, quand elles sont endommagées, sont associées aux performances à un test spécifique (Figure 4C). Cette méthode néanmoins ne permet pas de connaître les zones cérébrales aux alentours ou plus éloignées qui peuvent être touchées par la pathologie neurologique (lésion structurelle délimitée), notamment en ce qui concerne la connectivité anatomique (intégrité des faisceaux de la substance blanche), et la connectivité fonctionnelle (activité cérébrale).

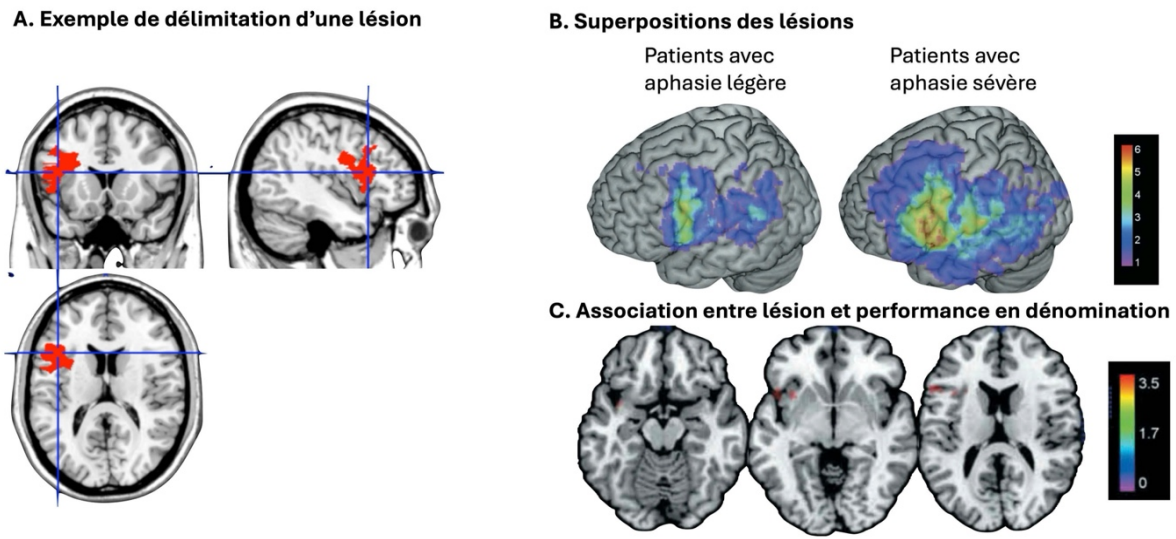

*Figure 4. Exemples d'analyses basées sur la délimitation des lésions (Døli et al., 2021). (A) Illustration d'une lésion vasculaire délimitée dans un espace de référence standard. (B) Superpositions des lésions de deux groupes de patient·es ayant subi un AVC gauche : à gauche, les patient·es avec peu d'atteinte langagière, et à droite, ceux présentant une aphasie sévère. On remarque que les patient·es avec une aphasie sévère présentent des lésions d'une plus grande étendue. (C) Exemple d'analyse en Voxel-Lesion Symptom Mapping (VLSM). Les zones colorées indiquent les régions cérébrales endommagées chez tous les patient·es, statistiquement associées aux performances à un test de dénomination.*

L'analyse morphométrique est une méthode qui permet d'identifier, à l'échelle du cerveau entier, les régions où l'on observe des différences de volumes (de matière grise et/ou blanche) entre deux groupes (ex. patient·es, sujets sains), ou de relier ces volumes à des performances cognitives. Cette méthode a l'avantage de pouvoir être utilisée aussi en cas de pathologie non vasculaire, par exemple pour mieux comprendre les associations anatomo-cliniques dans une population de patient·es présentant une maladie neurodégénérative.

L'IRM de diffusion permet d'étudier les faisceaux de substance blanche en quantifiant les caractéristiques microstructurales des tissus (axones, myéline) à travers le déplacement des molécules d'eau. Ces molécules subissent un mouvement aléatoire, appelé mouvement brownien, mais leur diffusion est influencée par la microstructure des tissus (protéines, myéline, filaments intracellulaires). Le liquide céphalo-rachidien est un milieu isotrope, c'est-à-dire un milieu dans lequel la diffusion des molécules d'eau va être uniforme, dans toutes les directions. En revanche, la substance blanche est un tissu anisotrope, ce qui signifie que ses propriétés varient selon l'orientation des fibres qui la composent. La diffusion des molécules d'eau va donc être limitée en raison de l'organisation parallèle des fibres créant une barrière biologique. Grâce à des algorithmes, l'IRM de diffusion permet de reconstruire des images en 3D représentant cette diffusion. Chaque petite région du cerveau (appelé voxel dans le cadre des images d'IRM) est caractérisée par un ellipsoïde, dont la forme varie selon la diffusion des molécules d'eau : une sphère pour un voxel isotrope (diffusion uniforme dans toutes les directions) et un ellipsoïde plus étiré pour un voxel anisotrope (direction orientée dans une direction). On peut déterminer la direction principale de la diffusion des molécules d'eau et extraire des valeurs permettant d'inférer l'intégrité des faisceaux de substance blanche dans chacun des voxels. À partir de ces données, il est ainsi possible de reconstruire les trajectoires des fibres de substance blanche de proche en proche (de voxel en voxel), en suivant la direction privilégiée définie par l'ellipsoïde. C'est ce que l'on appelle la tractographie (Figure 5). Elle s'effectue généralement en définissant des régions d'intérêt qui représentent des régions cérébrales dans lesquelles les faisceaux d'intérêt transitent (Catani & Thiebaut de Schotten, 2012).

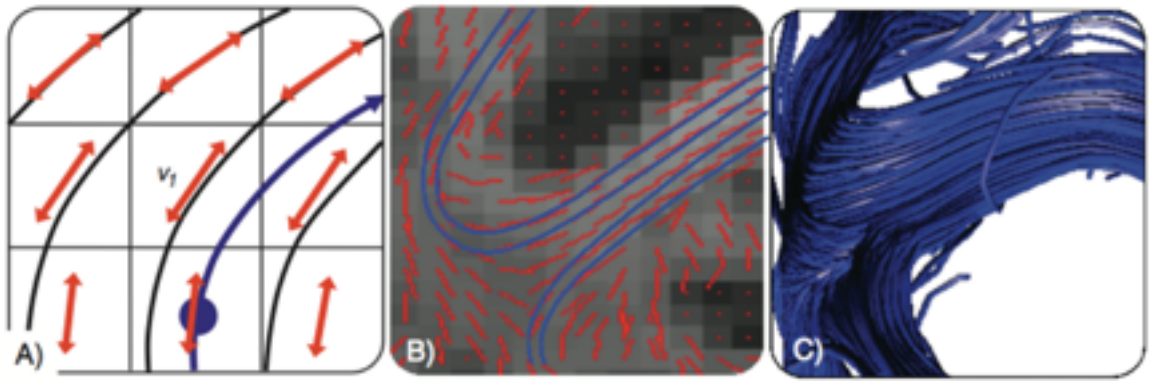

*Figure 5. Illustration de la méthode de tractographie des faisceaux de substance blanche. (A) les faisceaux sont reconstruits selon la direction principale de l'ellipsoïde de voxel en voxel. (B) Illustration des trajectoires des faisceaux reconstruits sur l'image obtenue en IRM de diffusion. (C) Illustration en 3D des faisceaux reconstruits.*

3.1.2   Imagerie fonctionnelle : l'Imagerie par Résonance Magnétique (IRM) fonctionnelle
L'IRM fonctionnelle enregistre les variations du flux sanguin dans le cerveau, qui sont liées à l'aimantation de l'hémoglobine contenue dans les globules rouges. Cette technique offre une excellente résolution spatiale, mais sa résolution temporelle est plus faible. Lors d'une activité cognitive, les régions cérébrales activées consomment plus d'oxygène, ce qui entraîne une augmentation du flux sanguin. Cela réduit la concentration de désoxyhémoglobine et augmente celle de l'oxyhémoglobine dans la région active concernée. Ce phénomène est mesuré par l'effet BOLD (Blood Oxygen Level Dependent), qui reflète les variations relatives entre ces deux formes d'hémoglobine. La désoxyhémoglobine, étant paramagnétique, réagit au champ magnétique de l'IRM en acquérant une aimantation détectable, ce qui permet de visualiser les régions d'activation cérébrale.

Afin de pouvoir identifier les régions cérébrales spécifiquement activées lors d'une tâche d'intérêt (ex. lecture de mots et pseudomots), il est essentiel de concevoir des protocoles expérimentaux très rigoureux. Généralement, on utilise des paradigmes dits « hiérarchisés » dans lesquels on enregistre l'activité cérébrale dans plusieurs conditions : celles des tâches d'intérêt (ex. lecture de mots et lecture de pseudomots), et une tâche contrôle (présentations de stimuli visuels non langagiers : barres obliques). Par la suite, on soustrait les images de la tâche contrôle de celle de la tâche d'intérêt pour mettre en évidence uniquement les activations cérébrales spécifiquement liées aux tâches expérimentales (Figure 6).

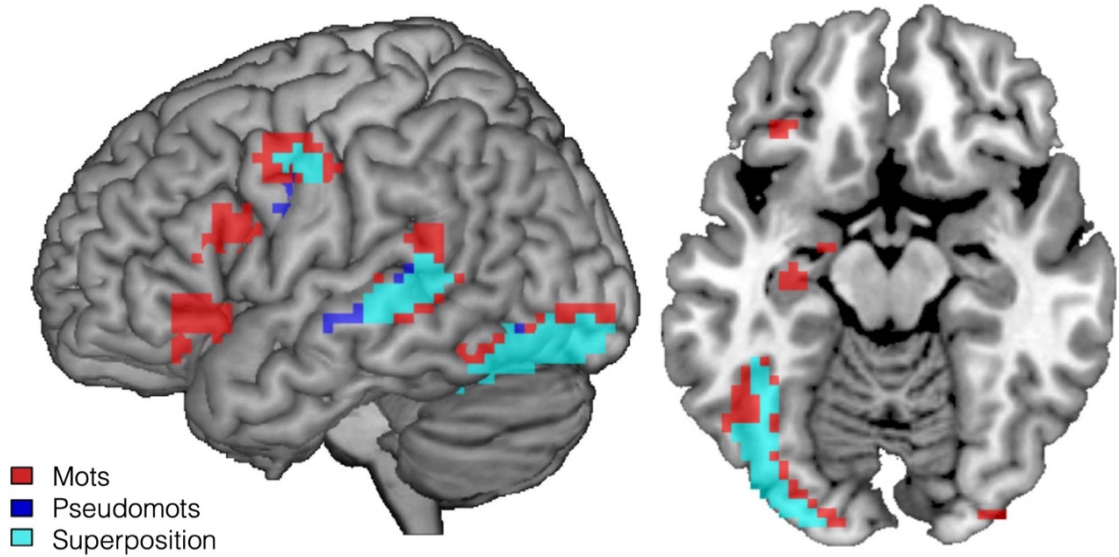

*Figure 6. Exemple d'étude en IRM fonctionnelle s'intéressant à identifier les régions cérébrales impliquées dans la lecture de mots et de pseudomots. La figure montre les activations pour les mots (en rouge) et les pseudomots (en bleu foncé) lorsqu'on contraste respectivement ces activations cérébrales avec celles obtenues lors de la présentation de stimuli visuels (barres obliques). La superposition des activations communes des mots et pseudomots apparait en bleu clair.*

### 3.2 Des zones cérébrales aux réseaux neuronaux

Dans le domaine du langage, ces méthodes d'imagerie cérébrale ont permis de confirmer le rôle prépondérant de l'hémisphère cérébral gauche pour les personnes droitières, cette latéralisation du langage dans l'hémisphère gauche étant parfois dans l'hémisphère droit chez les personnes gauchères ou moins latéralisées. Cependant ces méthodes ont aussi remis en question l'idée d'un centre de production dans la région de Broca et de compréhension dans la région de Wernicke. La différence entre parler et comprendre est nettement moins marquée que celle proposée par Broca et Wernicke. Même s'il est vrai que parler active majoritairement les aires frontales et comprendre majoritairement les aires temporo-pariétales, les deux régions sont impliquées à la fois dans la production et dans la compréhension de la parole. Nous savons désormais que c'est tout un réseau qui s'étend bien au-delà des aires de Broca et de Wernicke et qui touche également des régions sous-corticales – les ganglions de la base – cachées dans la profondeur du cerveau qui permet de parler et de comprendre le langage. Et, au-delà des régions corticales et des ganglions de la base, l'intégrité des connexions – la substance blanche – entre ces différentes régions, est tout aussi importante pour le traitement du langage (Roth et al., 2024).

Plus précisément, si l'on considère la région de Broca, de nombreuses études ont montré que cette région, considérée comme le centre de la parole, était aussi activée dans des tâches de compréhension de la parole remettant en question le dogme du centre de production de la parole (Flinker et al., 2015 ; Giraud, 2004 ; Papathanassiou et al., 2000). La région de Broca est activée également dans des tâches non langagières, comme la musique ou la mémoire visuelle (Maess et al., 2001 ; Mecklinger et al., 2004). Par ailleurs, une lésion de la région de Broca ne résulte pas toujours en une aphasie de Broca, et une aphasie de Broca n'implique pas nécessairement une lésion de l'aire de Broca (Dronkers, 1996 ; Dronkers et al., 2004 ; Gajardo-Vidal et al., 2021 ; Ochfeld et al., 2010). Enfin, le cerveau autopsié par Broca était un cerveau atrophié d'un patient qui souffrait de crises d'épilepsie depuis sa jeunesse et la localisation précise de l'aire de Broca au sein du gyrus frontal inférieur fait débat (Amunts & Zilles, 2006). Le modèle binaire classique Broca/Wernicke n'est ni un modèle linguistiquement correct, ni anatomo-fonctionnellement précis. L'utilisation de cette terminologie légitime artificiellement un modèle erroné malgré les données de neuroimagerie, de psycholinguistique et de pratique clinique des 30 dernières années. Pour ces multiples raisons, il est préférable, d'une part, de limiter l'utilisation du terme « région de Broca » et de spécifier plutôt les régions de manière anatomique (ex. gyrus frontal inférieur, pars opercularis), et, d'autre part, de reconsidérer complètement la neurolinguistique du traitement de la parole en intégrant les données récentes de la littérature médicale et scientifique (Poeppel et al., 2012 ; Poeppel & Hickok, 2004 ; Tremblay & Dick, 2016).

Les données d'imagerie permettent d'associer les différentes fonctions langagières à des réseaux cérébraux (Bambini et al., 2011 ; Bischetti et al., 2024 ; Crinion et al., 2006 ; Feng et al., 2021 ; Giavazzi et al., 2018 ; Hervais-Adelman et al., 2011 ; Hickok & Poeppel, 2007 ; Jacquemot et al., 2003 ; Jacquemot & Bachoud-Lévi, 2021b ; Lee & Dapretto, 2006 ; Mashal et al., 2007 ; Price, 2012 ; Witteman et al., 2011 ; voir les Figure 1 et Figure 2 et le Tableau 1 pour les localisations cérébrales impliquées dans les différentes composantes du traitement du langage).

La combinaison de tâches langagières spécifiques, de modèles lésionnels et de méthodes de neuro-imagerie a grandement contribué à mettre en évidence les liens structurels et fonctionnels entre différentes régions cérébrales et des capacités langagières précises. Par exemple, les tâches de dénomination d'objets sont couramment utilisées pour évaluer le traitement sémantique, la sélection lexicale ainsi que la planification et l'exécution motrice de l'articulation. Les études d'imagerie fonctionnelle chez les sujets sains montrent que la dénomination repose principalement sur l'activation du gyrus temporal moyen dans l'hémisphère gauche (Faria et al., 2005). Chez les patient·es atteint·es d'aphasie progressive ou de démence sémantique, l'atrophie de cette région a été corrélée aux déficits de dénomination. Toutefois, les altérations associées à ces déficits impliquent un réseau plus étendu, incluant une atrophie bilatérale des gyri temporaux inférieur et supérieur, des gyri fusiformes antérieurs et des hippocampes, ainsi qu'une atrophie unilatérale à gauche du gyrus parahippocampique, du gyrus temporal moyen et du pôle temporal (Brambati et al, 2006). Chez une large population de patient·es ayant subi un AVC de l'hémisphère gauche, des analyses en VLSM ont confirmé l'implication d'un large réseau cérébral. En effet, chez ces patient·es, les performances à des tests de dénomination comme le Boston Naming Test sont associées à un ensemble de structures comprenant les gyri temporaux moyen et supérieur, la substance blanche sous-jacente ainsi que le cortex pariétal inférieur (Baldo et al., 2013).

*Tableau 1 : Régions cérébrales impliquées dans les différentes composantes du traitement du langage.*

| Composantes langagières et communicationnelles | Régions cérébrales impliquées |
|---|---|
| Concept (ou sémantique) | - Zone antérieure des lobes temporaux (gyrus temporal moyen et inférieur) de façon bilatérale<br>- Hippocampes |
| Lexique | - Gyrus temporal moyen<br>- Jonction entre le gyrus temporal supérieur et le lobe pariétal<br>- Gyrus angulaire<br>- Gyrus frontal moyen et inférieur<br>- Ganglions de la base (noyau caudé) |
| Syntaxe | - Gyrus frontal inférieur (pars triangularis)<br>- Partie postérieure du cortex temporal moyen<br>- Ganglions de la base (putamen, noyau caudé) |
| Morphologie | - Gyrus frontal inférieur<br>- Gyrus temporal moyen<br>- Jonction entre le gyrus temporal supérieur et le lobule pariétal inférieur<br>- Ganglions de la base (putamen, noyau caudé) |
| Phonologie | - Gyrus temporal supérieur, notamment planum temporal, sulcus temporal supérieur et gyrus supramarginal, pour la compréhension<br>- Gyrus frontal inférieur, notamment la pars opercularis, et gyrus précentral, pour la production |
| Phonétique et traitement articulatoire | - Sillon frontal inférieur postérieur<br>- Aire motrice supplémentaire<br>- Cortex prémoteur ventral<br>- Ganglions de la base (putamen) |
| Prosodie linguistique | - Lobe frontal (bilatéralement)<br>- Gyrus temporal supérieur (bilatéralement)<br>- Gyrus temporal inférieur (bilatéralement) |
| Prosodie émotionnelle | - Lobe frontal (bilatéralement)<br>- Gyrus temporal supérieur (bilatéralement)<br>- Gyrus temporal inférieur (bilatéralement)<br>- Amygdale (bilatéralement) |
| Pragmatique | - Gyrus frontal inférieur (pars triangularis et opercularis)<br>- Gyrus précentral<br>- Gyrus temporal moyen<br>- Cortex cingulaire antérieur (bilatéralement)<br>- Lobule pariétal inférieur gauche, gyrus angulaire et sillon temporal supérieur droit, régions du réseau lié à la théorie de l'esprit (voir glossaire) |

En conclusion, ces données mettent en évidence le rôle des réseaux plutôt que l'implication de régions isolées dans les différentes composantes du langage : on parle de réseau cérébral. Les composantes sont associées à des régions corticales distantes, ce qui souligne l'importance de l'intégrité des fibres (substance blanche) qui vont connecter ces différentes régions cérébrales, notamment du faisceau arqué. Le fonctionnement du réseau dépend fortement des connexions qui relient les différents points du réseau et la lésion des connexions va entrainer des déficits tout aussi importants que la lésion des régions cérébrales elles-mêmes. Certaines régions sont associées à plusieurs composantes du langage : le gyrus frontal inférieur est impliqué dans le traitement de la phonologie et de la syntaxe entre autres ; et a contrario, une composante peut activer plusieurs régions : la phonologie est associée à l'activation du gyrus frontal inférieur, du gyrus temporal supérieur et du gyrus précentral, ce qui souligne l'importance de la dynamique des réseaux. La connectivité anatomique entre les différentes régions cérébrales est importante pour des capacités langagières efficientes, mais également pour un meilleur pronostic après une lésion cérébrale. Grâce à la méthode de tractographie de faisceaux de substance blanche, il a par exemple été montré que la récupération du langage à six mois post-AVC est associée au volume du faisceau arqué dans l'hémisphère droit (Forkel et al., 2014 ; Figure 7). Ce faisceau de substance blanche dans l'hémisphère gauche, bien connu pour son rôle dans les processus langagiers, relie les régions temporales, pariétales et frontales du langage. L'étude de Forkel et al. (2014) montre de façon très intéressante l'importance de son homologue dans l'hémisphère droit, qui semble jouer un rôle clé dans la récupération du langage après un AVC, lorsque le faisceau de l'hémisphère gauche est endommagé.

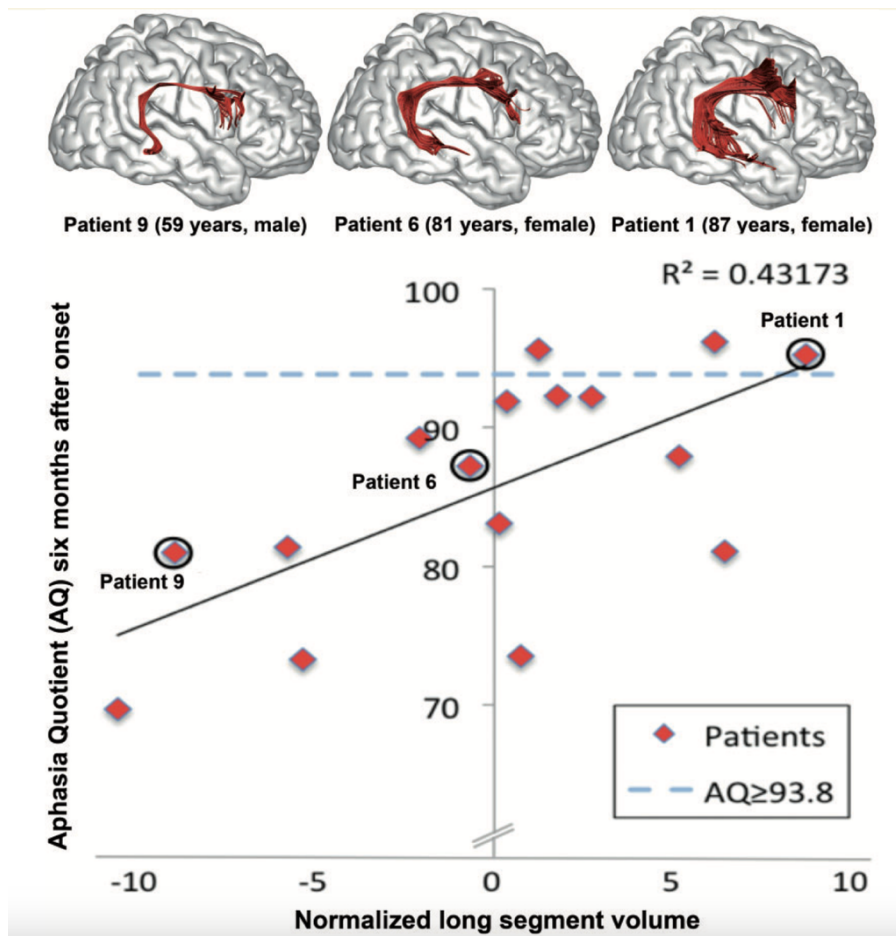

*Figure 7. Lien entre volume de substance blanche dans l'hémisphère droit et récupération de l'aphasie. Cette étude met en évidence une corrélation entre le volume du long segment du faisceau arqué et les scores à une batterie d'évaluation de l'aphasie, six mois après un AVC. Plus le volume du faisceau est important, meilleure est la récupération du langage. Le long segment du faisceau arqué connecte la région de Broca à celle de Wernicke et a été reconstruit dans les deux hémisphères par la méthode de tractographie en utilisant les données obtenues chez les patient·es en IRM de diffusion.*

## 3.3 Langage en interaction avec les autres fonctions cognitives

Les données des patient·es qui présentent des déficits spécifiques au langage ont permis de montrer que cette fonction peut être atteinte de manière ciblée. Cependant, les données cliniques sont souvent moins claires. Les déficits peuvent varier en fonction des conditions expérimentales et la plainte des patient·es peut ne pas être en accord avec les résultats aux tests. Cela s'explique par le fait que le langage, bien qu'étant une fonction à part entière, ne fonctionne pas indépendamment des autres fonctions. Comme on l'a vu, parler et comprendre nécessitent aussi des ressources attentionnelles, mnésiques et décisionnelles. Notamment de nombreuses tâches de langage, comme celle décrite au début de ce chapitre « Est-ce que tu peux me passer le sel ? » vont requérir également d'autres fonctions : la mémoire de travail, des ressources attentionnelles, des ressources exécutives, et la prise de décision. Les découvertes récentes de sciences cognitives ont montré que les déficits de mémoire de travail pouvaient avoir des répercussions importantes sur le traitement du langage, notamment pour comprendre des phrases ambiguës, alors que la compréhension des mots est intacte (Jacquemot et al., 2006 ; Jacquemot & Scott, 2006). Certaines lésions cérébrales, bien qu'elles n'affectent pas directement les régions classiquement associées au langage, peuvent néanmoins altérer les performances des patient·es dans des tâches linguistiques. Ces observations conduisent à repenser les réseaux langagiers en les intégrant à d'autres réseaux cognitifs. Par exemple, il a été montré que les patient·es souffrant d'une atrophie du striatum – une structure impliquée dans les processus cognitifs de prise de décision – présentent des déficits dans des tâches linguistiques (Jacquemot & Bachoud-Lévi, 2021a). L'analyse combinée des données comportementales et de neuroimagerie chez ces patient·es a permis de mettre en évidence l'impact d'un déficit des processus décisionnels sur le langage, enrichissant ainsi les connaissances sur le fonctionnement du langage en y intégrant d'autres fonctions cognitives (Le Stanc et al., 2023). L'attention et les ressources exécutives sont également essentielles pour réaliser les tâches de langage plus ou moins complexes, et des travaux récents qui couplent des études comportementales et des études d'imagerie cérébrale montrent que certaines ressources attentionnelles et exécutives sont partie intégrante du système de traitement du langage (Jacquemot & Bachoud-Lévi, 2021b, 2021a) ce qui pourrait expliquer certaines plaintes de patient·es difficiles à caractériser avec les tests de langage en pratique clinique. Ainsi, au réseau du langage à proprement parler, s'ajoutent les aires des réseaux de l'attention dans les régions pariétales, et de la mémoire de travail, dans les régions frontales inférieure et pariétale, et des ressources exécutives dans les régions frontales et des ganglions de la base.

## 4 Conclusion

L'étude des troubles linguistiques consécutifs à des lésions cérébrales a marqué un tournant décisif de la médecine au 19e siècle, ouvrant la voie à l'accumulation des connaissances sur le cerveau. L'analyse de cas de patient·es présentant des troubles de la compréhension ou de la production de la parole et du langage, ainsi que les liens établis avec certaines régions cérébrales, ont constitué une avancée majeure dans la compréhension des mécanismes du langage et des structures cérébrales impliquées. Ces dernières années, les techniques d'imagerie cérébrale ont franchi une nouvelle étape en offrant une vision plus intégrée du langage. Elles permettent non seulement d'identifier les régions cérébrales impliquées, mais aussi d'analyser les connexions qui les relient, en cohérence avec les découvertes récentes en psycholinguistique. Ces connaissances améliorent notre compréhension des troubles du langage et ouvrent la voie à de nouvelles approches de prise en charge et de rééducation.

# Glossaire

**Attention ou système attentionnel** : Capacité du système cognitif à se focaliser sur des stimuli spécifiques ou des tâches tout en ignorant d'autres distractions.

**Conduite d'approche** : Tentatives successives pour produire le mot cible, avec autocorrections et productions de variations phonémiques qui se rapprochent ou non de la cible (ex. *pes, pec, spec, petake, spectake, spectacle*).

**Déficit cognitif** : Un déficit cognitif désigne une altération des capacités mentales affectant des fonctions telles que la mémoire, l'attention, le langage, les fonctions exécutives (ex. inhibition, flexibilité, monitoring, planification), la prise de décision, le traitement visuel ou auditif, la capacité à traiter des concepts complexes.

**Déficit comportemental** : Un déficit comportemental désigne une altération des comportements adaptatifs dans les interactions sociales et de la régulation des émotions.

**Jargon :** Production de non-mots ou morceaux de phrases contenant des suites de mots dont aucun sens ne peut être extrait. Le jargon est souvent associé à une fluence excessive ou logorrhée.

**Mémoire épisodique** : Mémoire des expériences et des épisodes vécus. C'est la mémoire des événements de la vie personnelle.

**Mémoire explicite ou déclarative** : La mémoire sémantique et la mémoire épisodique constituent la mémoire explicite ou déclarative.

**Mémoire sémantique ou système conceptuel :** Stockage de toutes les connaissances générales acquises tout au long de la vie, s'appuyant sur les connaissances encyclopédiques et les connaissances conceptuelles sur le monde.

**Mémoire de travail** : La mémoire de travail est le système dédié au maintien des informations en mémoire à court terme (pendant quelques secondes) pour permettre leur traitement.

**Mot vide :** Mot sans signification précise (ex. truc, machin).

**Paraphasie** : Erreurs de production de la parole qui produit la transformation d'une unité par une autre.

- Paraphasie sémantique : production d'un mot sémantiquement proche du mot cible (ex. *fourchette* pour *cuillère*)
- Paraphasie phonémique : insertion, délétion ou substitution d'un ou plusieurs phonèmes (ex. *frenêtre* pour *fenêtre*, *bouton* pour *mouton*) donnant lieu à un mot ou un non-mot
- Paraphasie lexicale : production d'un mot sans lien avec le mot cible (ex. *balance* pour *stylo*)
- Paraphasie fantastique : production d'une périphrase qui ne partage aucun lien sémantique avec le mot cible (ex. *machine qui tient des relations épistolaires* pour *ordinateur*)

**Pause remplie :** Interruption dans le déroulement du discours marquée par l'insertion d'une unité sonore sans sens (ex. euh , hum).

**Périphrase ou circonlocution :** Erreur de substitution consistant à produire une expression ou une phrase plus longue que la cible, mais dont le sens est équivalent (ex. u*stensile que l'on met dans la serrure pour ouvrir la porte* pour *clef*)

**Prise de décision** : Processus cognitif aboutissant à la sélection d'une option parmi plusieurs options possibles

**Théorie de l'esprit :** La théorie de l'esprit est un concept qui désigne la capacité d'un individu à attribuer des états mentaux – comme des croyances, des désirs, des intentions ou des émotions – à soi-même et aux autres. Cette capacité permet de comprendre que les autres ont des pensées, des sentiments, et des perspectives différentes des nôtres. Ce système joue un rôle crucial dans les interactions sociales, car il permet de comprendre, de prédire et d'anticiper les réactions des autres, facilitant ainsi la communication et la coopération.

# Références

Amunts, K., & Zilles, K. (2006). A multimodal analysis of structure and function in Broca's region. In Y. Grodzinsky & K. Amunts (Eds.), *Broca's region* (p. 17-30). Oxford University Press.

Bachoud-Lévi, A.-C., Dormeuil, A., & Jacquemot, C. (2022). Improving efficacy of aphasia rehabilitation by using Core Assessment of Language Processing. *Annals of Physical and Rehabilitation Medicine*, *65*(6), 101630.

Baldo, J. V., Arévalo, A., Patterson, J. P., & Dronkers, N. F. (2013). Grey and white matter correlates of picture naming: evidence from a voxel-based lesion analysis of the Boston Naming Test. *Cortex*, 49(3), 658-667.

Bambini, V., Gentili, C., Ricciardi, E., Bertinetto, P. M., & Pietrini, P. (2011). Decomposing metaphor processing at the cognitive and neural level through functional magnetic resonance imaging. *Brain Research Bulletin*, *86*(3–4), 203–216.

Bischetti, L., Frau, F., & Bambini, V. (2024). Neuropragmatics. In M. J. Ball, N. Müller, & E. Spencer (Eds.), *The handbook of clinical linguistics. Second Edition* (p. 41–54). Wiley.

Brambati, S. M., Myers, D., Wilson, A., Rankin, K. P., Allison, S. C., Rosen, H. J., ... & Gorno-Tempini, M. L. (2006). The anatomy of category-specific object naming in neurodegenerative diseases. *Journal of Cognitive Neuroscience*, *18*(10), 1644-1653.

Caramazza, A. (1997). How many levels of processing are there in lexical access? *Cognitive Neuropsychology*, *14*(1), 177-208.

Catani, M., & Thiebaut de Schotten, M. (2012). *Atlas of human brain connections*. Oxford University Press.

Crinion, J., Turner, R., Grogan, A., Hanakawa, T., Noppeney, U., Devlin, J. T., Aso, T., Urayama, S., Fukuyama, H., Stockton, K., Usui, K., Green, D. W., & Price, C. J. (2006). Language control in the bilingual brain. *Science*, *312*(5779), 1537–1540.

Dell, G. S., & O'Seaghdha, P. G. (1992). Stages of lexical access in language production. *Cognition*, *42*(1–3), 287–314.

Døli, H., Andersen Helland, W., Helland, T., & Specht, K. (2021). Associations between lesion size, lesion location and aphasia in acute stroke. *Aphasiology*, *35*(6), 745–763.

Dronkers, N. F. (1996). A new brain region for coordinating speech articulation. *Nature*, *384*(6605), 159–161.

Dronkers, N. F., Wilkins, D. P., Van Valin, R. D., Redfern, B. B., & Jaeger, J. J. (2004). Lesion analysis of the brain areas involved in language comprehension. *Cognition*, *92*(1–2), 145–177.

Farias, S. T., Harrington, G., Broomand, C., & Seyal, M. (2005). Differences in functional MR imaging activation patterns associated with confrontation naming and responsive naming. *American Journal of Neuroradiology*, *26*(10), 2492-2499.

Feng, W., Wang, W., Liu, J., Wang, Z., Tian, L., & Fan, L. (2021). Neural correlates of causal inferences in discourse understanding and logical problem-solving: A meta-analysis study. *Frontiers in Human Neuroscience*, *15*, 666179.

Flinker, A., Korzeniewska, A., Shestyuk, A. Y., Franaszczuk, P. J., Dronkers, N. F., Knight, R. T., & Crone, N. E. (2015). Redefining the role of Broca's area in speech. *Proceedings of the National Academy of Sciences*, *112*(9), 2871–2875.

Forkel, S. J., Thiebaut De Schotten, M., Dell'Acqua, F., Kalra, L., Murphy, D. G. M., Williams, S. C. R., & Catani, M. (2014). Anatomical predictors of aphasia recovery: A tractography study of bilateral perisylvian language networks. *Brain*, *137*(7), 2027–2039.

Franklin, S. (1989). Dissociations in auditory word comprehension; evidence from nine fluent aphasic patients. *Aphasiology*, *3*(3), 189–207.

Friederici, A. D. (2011). The brain basis of language processing: From structure to function. *Physiological Reviews*, *91*(4), 1357–1392.

Gajardo-Vidal, A., Lorca-Puls, D. L., Team, P., Warner, H., Pshdary, B., Crinion, J. T., Leff, A. P., Hope, T. M. H., Geva, S., Seghier, M. L., Green, D. W., Bowman, H., & Price, C. J. (2021). Damage to Broca's area does not contribute to long-term speech production outcome after stroke. *Brain*, *144*(3), 817–832.

Gatignol, P., Jutteau, S., Oudry, M., & Weill-Chounlamountry, A. (2012). *Bilan Informatisé d'Aphasie*. Ortho Edition.

Giavazzi, M., Daland, R., Peperkamp, S., Palminteri, S., Brugières, P., Jacquemot, C., Schramm, C., Cleret de Langavant, L., Bachoud-Lévi, A.-C., Peperkamp, S., Brugieres, P., Jacquemot, C., Schramm, C., Cleret de Langavant, L., & Bachoud-Levi, A. C. (2018). The role of the striatum in linguistic selection: Evidence from Huntington's disease and computational modeling. *Cortex*, *109*, 189–204.

Giraud, A. L. (2004). Contributions of sensory input, auditory search and verbal comprehension to cortical activity during speech processing. *Cerebral Cortex*, *14*(3), 247–255.

Hervais-Adelman, A. G., Moser-Mercer, B., & Golestani, N. (2011). Executive control of language in the bilingual brain: Integrating the evidence from neuroimaging to neuropsychology. *Frontiers in Psychology*, *2*, 234–234.

Hickok, G., & Poeppel, D. (2007). The cortical organization of speech processing. *Nature Reviews Neuroscience*, *8*(5), 393–402.

Hillis, A. (1993). The role of models of language processing in rehabilitation of language impairments. *Aphasiology*, *7*(1), 5–26.

Hillis, A. (1998). Treatment of naming disorders: New issues regarding old therapies. *Journal of the International Neuropsychological Society*, *4*(6), 648–660.

Hillis, A., & Caramazza, A. (1994). Theories of lexical processing and rehabilitation of lexical deficits. In M. J. Riddoch & G. E. Humphreys (Eds.), *Cognitive neuropsychology and cognitive rehabilitation*. Lawrence Erlbaum Associates.

Jacquemot, C., & Bachoud-Lévi, A.-C. (2021a). A case-study of language-specific executive disorder. *Cognitive Neuropsychology*, *38*(2), 125-137.

Jacquemot, C., & Bachoud-Lévi, A.-C. (2021b). Striatum and language processing: Where do we stand? *Cognition*, *213*, 104785.

Jacquemot, C., Dupoux, E., & Bachoud-Levi, A. C. (2011). Is the word-length effect linked to subvocal rehearsal? *Cortex*, *47*(4), 484–493.

Jacquemot, C., Dupoux, E., Decouche, O., & Bachoud-Levi, A.-C. (2006). Misperception in sentences but not in words: Speech perception and the phonological buffer. *Cognitive Neuropsychology*, *23*(6), 949-971.

Jacquemot, C., Dupoux, E., Robotham, L., & Bachoud-Levi, A. C. (2012). Specificity in rehabilitation of word production: A meta-analysis and a case study. *Behavioural Neurology*, *25*(2), 73–101.

Jacquemot, C., Lalanne, C., Sliwinski, A., Piccinini, P., Dupoux, E., & Bachoud-Levi, A. C. (2019). Improving language evaluation in neurological disorders: The French Core Assessment of Language Processing (CALAP). *Psychological Assessment*, *31*(5), 622-630.

Jacquemot, C., Pallier, C., LeBihan, D., Dehaene, S., & Dupoux, E. (2003). Phonological grammar shapes the auditory cortex: A functional magnetic resonance imaging study. *Journal of Neuroscience*, *23*(29), 9541–9546.

Jacquemot, C., & Scott, S. K. (2006). What is the relationship between phonological short-term memory and speech processing? *Trends in Cognitive Sciences*, *10*(11), 480–486.

Le Stanc, L., Youssov, K., Giavazzi, M., Sliwinski, A., Bachoud-Lévi, A. C., & Jacquemot, C. (2023). Language disorders in patients with striatal lesions: Deciphering the role of the striatum in language performance. *Cortex*, *166*, 91-106.

Lee, S. S., & Dapretto, M. (2006). Metaphorical vs. literal word meanings: fMRI evidence against a selective role of the right hemisphere. *NeuroImage*, *29*(2), 536–544.

Levelt, W. J. M., Roelofs, A. P. A., & Meyer, A. S. (1999). A theory of lexical access in speech production. *Behavioral and Brain Sciences*, *22*(1), 1–75.

Maess, B., Koelsch, S., Gunter, T. C., & Friederici, A. D. (2001). Musical syntax is processed in Broca's area: An MEG study. *Nature Neuroscience*, *4*(5), 540–545.

Mashal, N., Faust, M., Hendler, T., & Jung-Beeman, M. (2007). An fMRI investigation of the neural correlates underlying the processing of novel metaphoric expressions. *Brain and Language*, *100*(2), 115–126.

Mecklinger, A., Gruenewald, C., Weiskopf, N., & Doeller, C. F. (2004). Motor affordance and its role for visual working memory: Evidence from fMRI studies. *Experimental Psychology*, *51*(4), 258–269.

Miceli, G., Capasso, R., & Caramazza, A. (1994). The interaction of lexical and sublexical processes in reading, writing and repetition. *Neuropsychologia*, *32*(3), 317–333.

Nespoulous, J.-L., Roch Lecours, A., Lafond, D., Lemay, A., Puel, M., Joanette, Y., Cot, F., & Rascol, A. (1992). MT 86. Protocole Montréal-Toulouse d'examen linguistique de l'aphasie. Ortho Edition.

Nickels, L. (2002). Therapy for naming disorders: Revisiting, revising and reviewing. *Aphasiology*, *16*(10–11), 935–980.

Ochfeld, E., Newhart, M., Molitoris, J., Leigh, R., Cloutman, L., Davis, C., Crinion, J., & Hillis, A. E. (2010). Ischemia in broca area is associated with broca aphasia more reliably in acute than in chronic stroke. *Stroke*, *41*(2), 325–330.

Papathanassiou, D., Etard, O., Mellet, E., Zago, L., Mazoyer, B., & Tzourio-Mazoyer, N. (2000). A common language network for comprehension and production: A contribution to the definition of language epicenters with PET. *NeuroImage*, *11*(4), 347–357.

Poeppel, D., Emmorey, K., Hickok, G., & Pylkkanen, L. (2012). Towards a new neurobiology of language. *Journal of Neuroscience*, *32*(41), 14125–14131.

Poeppel, D., & Hickok, G. (2004). Towards a new functional anatomy of language. *Cognition*, *92*(1–2), 1–12.

Price, C. J. (2012). A review and synthesis of the first 20 years of PET and fMRI studies of heard speech, spoken language and reading. *Neuroimage*, *62*(2), 816–847.

Ramoo, D., Olson, A., & Romani, C. (2021). Repeated attempts, phonetic errors, and syllabifications in a case study:Evidence of impaired transfer from phonology to articulatory planning. *Aphasiology*, *35*(4), 485–517.

Roth, R. W., Schwen Blackett, D., Gleichgerrcht, E., Wilmskoetter, J., Rorden, C., Newman-Norlund, R., Sen, S., Fridriksson, J., Busby, N., & Bonilha, L. (2024). Long-range white matter fibers and post-stroke verbal and non-verbal cognition. *Brain Communications*, *6*(4), fcae262.

Tran, T. M., & Godefroy, O. (2011). La Batterie d'Évaluation des Troubles Lexicaux: Effet des variables démographiques et linguistiques, reproductibilité et seuils préliminaires. *Revue de Neuropsychologie*, *3*(1), 52–69.

Tremblay, P., & Dick, A. S. (2016). Broca and Wernicke are dead, or moving past the classic model of language neurobiology. *Brain and Language*, *162*, 60–71.

Witteman, J., Van IJzendoorn, M. H., Van De Velde, D., Van Heuven, V. J. J. P., & Schiller, N. O. (2011). The nature of hemispheric specialization for linguistic and emotional prosodic perception: A meta-analysis of the lesion literature. *Neuropsychologia*, *49*(13), 3722–3738.